# Atomistic full-band simulations of monolayer MoS$_2$ transistors

Jiwon Chang, Leonard F. Register and Sanjay K. Banerjee

Microelectronics Research Center, The University of Texas at Austin, Austin, TX 78758, USA

We study the transport properties of deeply scaled monolayer MoS$_2$ n-channel metal-oxide-semiconductor field effect transistors (MOSFETs) using full-band ballistic quantum transport simulations with an atomistic tight-binding Hamiltonian obtained from density functional theory. Our simulations suggest that monolayer MoS$_2$ MOSFETs can provide near-ideal subthreshold slope, and suppression of drain-induced barrier lowering (DIBL) and gate-induced drain leakage (GIDL). However, these full-band simulations also exhibit limited transconductance. These ballistic simulations also exhibit negative differential resistance (NDR) in the output characteristics associated with the narrow width in energy of the lowest conduction band, but this NDR may be substantially reduced or eliminated by scattering in MoS$_2$.

Quasi-two-dimensional (2-D) layered materials have attracted much attention since the realization of isolated graphene monolayers [1]. Molybdenum disulfide ($MoS_2$) belongs to the family of 2-D-layered transition metal dichalcogenides (TMDs). A monolayer of $MoS_2$ is formed by hexagonally arranged atomic sheets of Mo and S atoms stacked together in an S–Mo–S sandwich. Two adjacent $MoS_2$ monolayers are weakly bonded by Van-der-Waals forces, allowing exfoliation of monolayers from bulk $MoS_2$ [2,3]. Due to their atomic scale thickness, monolayers of $MoS_2$ offer a high degree of electrostatic gate control, making them a promising material for low voltage switching. Recently, a monolayer [4] and multilayer $MoS_2$ [5,6] transistors have been realized with high mobility, high ON-OFF current ratio and good subthreshold slope. Integrated circuits based on bilayer [7] and monolayer $MoS_2$ [8] have been demonstrated as well. Ballistic, effective-mass-based non-equilibrium Green's function (NEGF) simulations of a 15 nm channel length monolayer $MoS_2$ MOSFET support the potential of such nano-scale devices [9,10]. A mean free path of 15 to 22 nm is also suggested in [9], based in part on the experimental work of [4], such that at least quasi-ballistic transport might be expected on this scale. To explore the limits of device performance further, we have performed ballistic *full-band* quantum transport simulations for a similar 15 nm channel length monolayer $MoS_2$ MOSFET. We find excellent subthreshold slope and limited short channel effects but more limited transconductance compared to [10]. Also, as a result of the relatively narrow in energy first conduction bands (CB), a region of negative differential resistance (NDR) in the drain current $I_{DS}$ vs. drain voltage $V_{DS}$ is found. However, depending on the scattering strength—which would depend on the quality of the $MoS_2$ monolayer and its environment—the NDR may be substantially reduced or eliminated.

The simulated n-channel transistor is illustrated in Fig. 1(b). Monolayer MoS$_2$ modeled with a dielectric constant $\kappa$ of 3.0 is placed on top of 50 nm of SiO$_2$ with $\kappa$ of 3.9. The undoped 15 nm channel is gated through 2.8 nm of HfO$_2$ with $\kappa$ of 25, as in [9, 10]. The in-plane $\kappa$ for monolayer MoS$_2$ has been estimated to be 3.3 [11], but the error is small and, with only a monolayer, the surrounding materials should dominate. The source and drain sheet donor doping is $3.5 \times 10^{13}$ cm$^{-2}$ with a corresponding Fermi level ~50 meV above the CB edge.

The primitive unit cell of monolayer MoS$_2$ is hexagonal with a three-atom basis, as indicated by the rhombus in Fig. 1(a). The band structure of the hexagonal Brillouin zone (BZ) (Fig. 1(c)) was calculated by the density functional theory (DFT) code OPENMX [12] using a Troullier and Martins (TM) type norm-conserving pseudopotential [13] with a partial core correction and a linear combination of pseudoatomic orbitals [14] as a basis set. Pseudoatomic orbitals were generated by a confinement scheme [14] with the cutoff radius 5.0 a.u. and 7.0 a.u. for Mo and S, respectively. Basis sets with s2p2d1 and s1p1d1 for Mo and S, respectively, were found to be good enough to reproduce the band structure obtained by plane wave based DFT calculations [15,16,17]. We adopted the local density approximation (LDA) [18] for exchange-correlation energy functional and used a kinetic energy cutoff of 200 Ryd and $k$-mesh size of 7×7×1. Since it has been reported that DFT calculations employing the experimental lattice constants reproduce the band gap of monolayer MoS$_2$ well [15, 17], we constructed the monolayer MoS$_2$ structure with experimental lattice parameters (Fig. 1(a)) [19] in our calculations. As with the previous studies [15,16,17,20], we found a direct gap with band edges at the K point. The calculated band gap of 1.8 eV is close to the experimentally measured value [2]. The CB effective mass is about 0.55 times the electron free-space mass, similar to the previous DFT calculation [17]. While results from other DFT calculations may slightly differ even with the same methods, there is consensus about the salient features of narrow energy dispersion of the conduction bands and correspondingly large effective masses.

To model ballistic quantum transport (partially considering reflection within the simulation region), propagating wave functions were obtained using a recursive scattering matrix approach [21] with perfectly absorbing boundaries. Moreover, all transport calculations were performed within an atomistic tight-binding (TB) basis of maximally localized Wannier functions (MLWFs) orbitals [22], with five centered about each Mo atom, and four centered about each S atom. The Wannier functions and onsite through 3$^{rd}$ nearest neighbor hopping potentials were calculated directly from the DFT Kohn-Sham orbitals and potential also using OPENMX. The resulting TB band structure reproduced well the original DFT band structure (LHS of Fig. 1(c)). To this end, we employed rectangular unit cells with a six-atom basis (rectangle (in orange online) in Fig. 1(a)) with an associated reduced Brillouin zone (Fig. 1(d)). The incident plane waves were resolved with uniform energy spacing $\Delta E < 2$ meV, and $N_y = 200$ uniformly separated values of $k_y$ were used to keep the associated energy spacing in the first conduction band small (the maximum energy spacing of ~6 meV). The propagating wavefunctions were normalized to an *incident* current density of $e(\Delta E/\pi\hbar)$ per incident mode per unit device width $\Delta W = N_y a_y$, assuming spin degeneracy within the conduction band, consistent with Landauer-Büttiker theory. Total charge density and *transmitted* current were obtained by summing over these contributions, weighted by the Fermi distributions of the injecting leads. The transport equations were solved self-consistently with Poisson's equation. All simulations were performed at 300 K.

Simulation results are shown in Fig. 2. With the carriers confined in the MoS$_2$ monolayer, this 15 nm planar device still exhibits excellent electrostatics. The transfer characteristics, drain current $I_{DS}$ vs. gate voltage $V_{GS}$ minus threshold voltage $V_T$, (Fig. 2(a)) exhibit limited drain-induced barrier lowering (DIBL) of ~10 mV/V, and a near-ideal subthreshold slope of ~60 mV/decade at 300 K. Gate-induced drain leakage (GIDL) is eliminated within the considered voltage ranges by the large band gap and resulting lack of CB-to-VB (valence band) overlap. The peak transconductance, reached near $V_{DS} = 0.2$ V, is about 4 mA/μm/V.

To benchmark our results with prior work, we intentionally chose some similar device parameters, such as identical gate stacks and channel length [9,10]. However, in [9], the authors assumed Schottky barrier contacts to the channel unlike us, which makes one-to-one comparisons difficult. However, our subthreshold characteristics such as a subthreshold slope and DIBL for which the effect of a Schottky barrier is limited, are very similar to [9]. In [10], where contacts appear to be more ideal and saturation is also reached in the 0.2 to 0.3 V range, a somewhat larger transconductance, ~6 mA/μm/V, is obtained, despite slightly heavier band-edge effective masses. We can speculate that the reduced transconductance here is due to some combination of reduced channel quantum (density of states) capacitance via fewer band-edge valleys than were obtained in [10] from different band structure calculations, and overall reduced carrier velocities via the combination of substantial non-parabolicity and degenerate carrier concentrations, perhaps exacerbated by still slower down-channel electrons through self-consistent electrostatics. (We note the gate stacks, channel length and substrate are identical to [9,10], except perhaps in the detailed treatment of the $MoS_2$ layer, and, thus, are not likely the source of these differences.)

Increasing $V_{DS}$ further, above about 0.2 V, in these full-band ballistic simulations then results in the previously mentioned NDR with respect to the $V_{DS}$, as seen in Fig. 2(b), and reduces the transconductance below 1.6 mA/μm/V with $V_{DS} = $ 0.5 V, as seen in Fig. 2(a). However, given that the NDR can be expected to be severely degraded if not eliminated by scattering as discussed below, we do not wish to overemphasize its practical importance here. However, this NDR is an unavoidable result of the combination of ballistic full band simulations and narrow energy dispersion of the band structure of $MoS_2$, and this will be clarified below.

While the barrier to electron injection from the source to channel is defined by bottom of the conduction band, for $MoS_2$ we must also consider the top or local maxima of the conduction band, since it can present a barrier to capture of ballistic electrons in the drain for moderate drain voltages. To illustrate, we consider transmission probabilities as a function of the voltage difference between the source and drain in Fig 3. In contrast to the electrostatically self-consistent device simulations of Fig. 2, here we use an artificial piecewise linear potential approximation for the source, channel and drain as a function of $V_{DS}$ (inset on the right-hand-sides (RHS) of Fig. 3(a)) to isolate the essential physics. Even for fixed energy and $k_y$, there are multiple transmission probabilities to consider between incident modes in the source and outgoing mode in the drain. The left-hand-sides (LHS) of Fig. 3(a)-(d) show the sum over these transmission probabilities as a function of energy for $k_y = 0$. The RHS provides the transmission probabilities between the indicated individual source and drain modes of the same spin at $k_y = 0$ and, specifically, the Fermi level (except in Fig. 3(a) with no potential drop and unity intra-mode and zero inter-mode transmission probabilities as required). The various symbols on the displayed source and drain lead band structures—the latter differing only in band-edge energy—on the RHS identify these modes. Note that in all cases there are four incoming states from the source lead including spin degeneracy, all in the lowest lying conduction band. There is a corresponding sum of transmission probabilities of essentially four up through $V_{DS} = 0.2$ V. However, just before $V_{DS}$ reaches 0.3 V, the number of available outgoing modes in the same (reduced rectangular) band in the drain which can be reached *semiclassically* is reduce to two, including spin degeneracy, and then is reduced again to zero near $V_{DS}$ = 0.4 V. The sum of transmission probabilities drops to about 2.5 and just below unity at these respective voltages. That these total transmission probabilities do not drop to $\leq$ 2 and 0, respectively, is due to tunneling. (In term of the rectangular bands structure of Fig 1(d), this tunneling is to states in the next highest band. However, in terms of the hexagonal band structure of Fig. 1(c), the tunneling is to states within the same band at least initially, but to ones that cannot be reached from the incoming modes semiclassically.) Of course, by comparison, the MOSFET $I_{DS}$-$V_{DS}$ behavior of Fig. 2 is smoothed by integration over energy and $k_y$, and an electrostatically self-consistent and more complicated potential geometry, but the onset of strong NDR occurs around $V_{DS}$ = 0.4 V, consistent with the above discussion. Scattering, however, not considered here would reduce or eliminate the NDR by, specifically, creating inelastic semiclassical intra-band and perhaps inter-band pathways into the drain at high $V_{DS}$.

In summary and conclusion, we used atomistic full-band ballistic quantum transport simulations with TB potentials obtained from DFT, to investigate the characteristics of 15 nm channel length monolayer $MoS_2$ MOSFETs. The salient features of this system captured in these simulations include a near-atomically thin channel, a large band gap, a high band-edge effective mass exacerbated by non-parabolicity, and a narrow-energy-dispersion first CB. Results exhibit an almost ideal subthreshold slope, small DIBL and suppression of GIDL. However, they also exhibit more limited transconductance than an effective mass model predicts. Our ballistic simulations also exhibit NDR in the $I_{DS}$ vs. $V_{DS}$ characteristics associated with the narrow first CB. Such NDR can be expected to be substantially reduced or eliminated by scattering which would depend on both the $MoS_2$ monolayer and its environment. Conversely, any observed NDR in nano-scale MoS$_2$ MOSFETs could be a signature of limited quasi-ballistic transport. In any case, NDR in ballistic full-band simulation of MOSFETs based on MoS$_2$ is to be expected and must be considered when interpreting results.

The authors acknowledge support from the Nanoelectronics Research Initiative supported Southwest Academy of Nanoelectronics (NRI-SWAN), Intel and NSF NASCENT ERC.

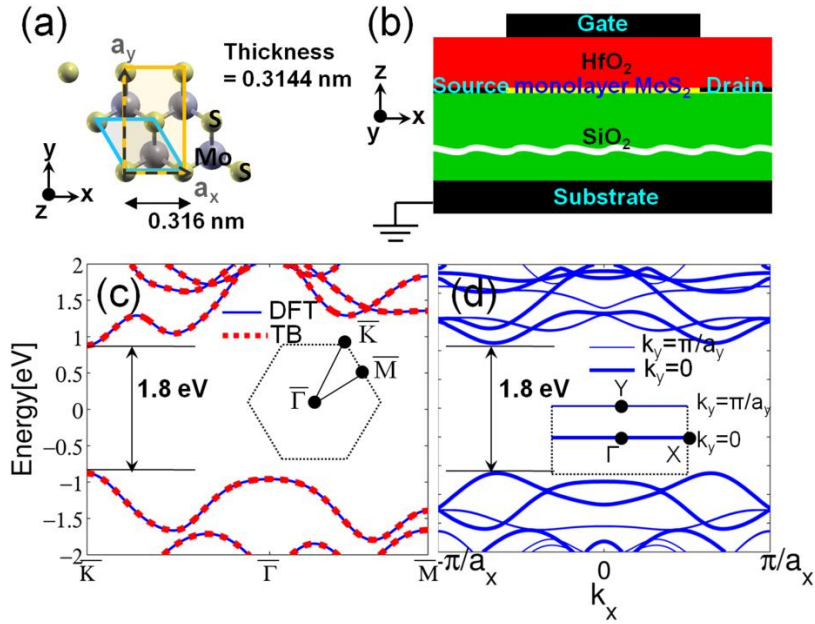

**FIG. 1**. (a) Top view of monolayer MoS$_2$ showing the hexagonal (light blue) and rectangular (orange) unit cells. (b) Device structure of monolayer MoS$_2$ transistor. The nominal device parameters are as follows: HfO$_2$ ($\kappa$ = 25) gate oxide thickness = 2.8 nm, channel length = 15 nm, n-type doping density of source and drain = $3.5 \times 10^{13}$ cm$^{-2}$, and SiO$_2$ oxide thickness = 50 nm. Band structures of monolayer MoS$_2$ (c) from DFT and TB Hamiltonian in the hexagonal Brillouin Zone (BZ) and (d) from the TB Hamiltonian at two different transverse modes ($k_y$ = 0 and $\pi/a_y$) in the rectangular BZ.

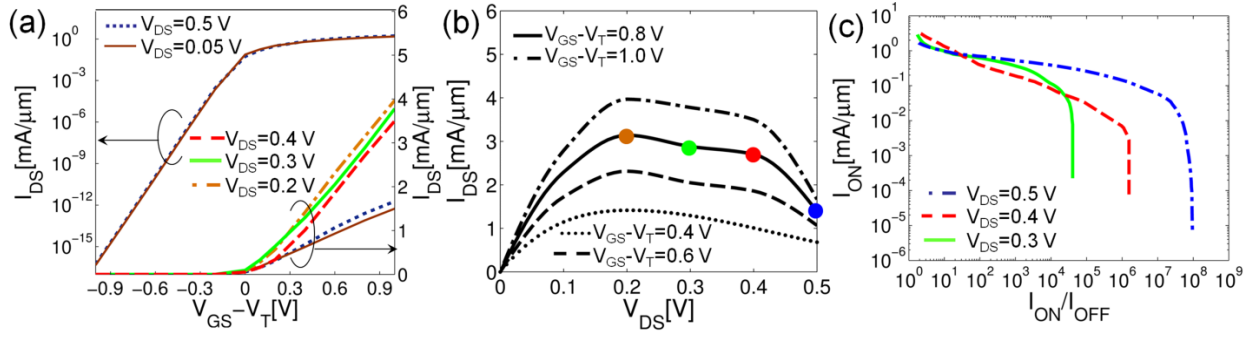

**FIG. 2**. Monolayer MoS$_2$ transistor characteristics: (a) $I_{DS}$ vs. $V_{GS}-V_T$ curves at $V_{DS}$ = 0.05 and 0.5 V on log scale, and $V_{DS}$ = 0.05, 0.2, 0.3, 0.4, and 0.5 V on linear scale. (b) $I_{DS}$ vs. $V_{DS}$ curves at $V_{GS}-V_T$ = 0.4, 0.6, 0.8 and 1.0 V; dots correspond to $V_{DS}$ in (a). (c) $I_{ON}$ vs. $I_{ON}/I_{OFF}$ at $V_{ON}-V_{OFF}$ = $V_{DS}$ = 0.3, 0.4 and 0.5 V.

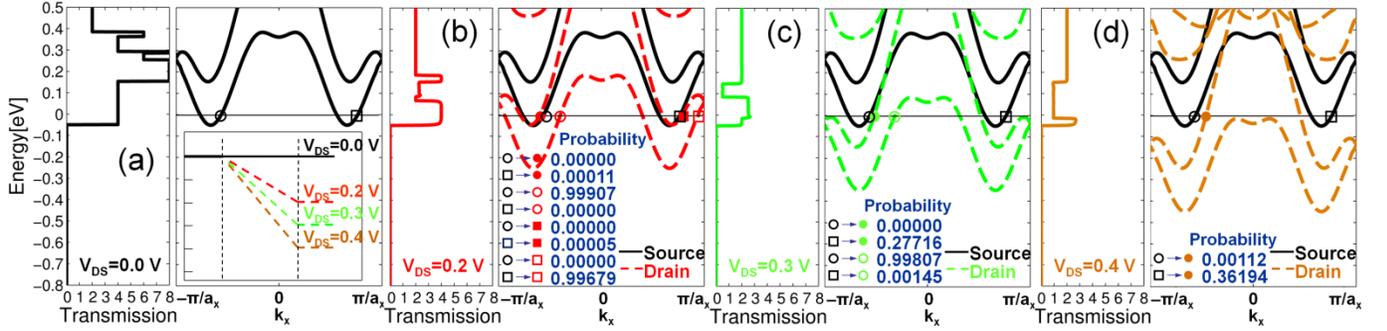

**FIG. 3**. Transmission probabilities between source and drain modes for drain biases of (a) $V_{DS}$ = 0.0 V, (b) $V_{DS}$ = 0.2 V, (c) $V_{DS}$ = 0.3 V and (d) $V_{DS}$ = 0.4 V, respectively, obtained using the piecewise constant potentials shown in the inset on the RHS of Fig. 1(a). The LHSs show the sum over all transmission probabilities for $k_y$ = 0, as a function of energy. The symbols on the source and drain band structures—differing only in band-edge energy—on the RHS graphically show those incoming and outgoing modes at the source Fermi level and $k_y$ = 0; the text below indicates the corresponding intra- and inter-mode transmission probabilities between available modes of the same spin, except in (a) where the solution is trivial.